\documentclass[a4paper,11pt]{article}
\usepackage[latin1]{inputenc}
\usepackage[T1]{fontenc}
\usepackage[english]{babel}
\usepackage{indentfirst}
\usepackage{amsmath,amssymb}
\usepackage{dsfont}
\usepackage{geometry}
\usepackage{subfig}
\usepackage{enumitem}
\usepackage{amsthm}
\usepackage{graphicx}
\usepackage{float}
\usepackage{mathtools}

\newtheorem{remark}{Remark}

\newtheorem{definition}{Definition}
\newtheorem{assumption}{Assumption}
\newtheorem{example}{Example}[section]

\usepackage{hyperref}
\begin{document}

\title{Nonparametric and arbitrage-free construction of call surfaces using $l_1$-recovery}
\author{P. M. Blacque-Florentin\footnote{Department of Mathematics, Imperial College, London SW7 2AZ, pb910@ic.ac.uk}, B. Missaoui\footnote{Department of Mathematics, Imperial College, London SW7 2AZ, badr.missaoui@imperial.ac.uk}}
\date{}
\maketitle

\begin{center}
\textbf{Abstract}
This paper is devoted to the application of an $l_1$ -minimisation technique to
construct an arbitrage-free call-option surface. We propose a nononparametric approach to
obtaining model-free call option surfaces that are perfectly consistent with market quotes
and free of static arbitrage. The approach is inspired from the compressed-sensing
framework that is used in signal processing to deal with under-sampled signals. We
address the problem of fitting the call-option surface to sparse option data. To illustrate the methodology, we proceed to the construction of the whole call-price surface of the S\&P500 options, taking into account the arbitrage possibilities in the time direction. The resulting object is a surface free of both butterfly and calendar-spread arbitrage that matches the original market points. We then move on to an FX application, namely the HKD/USD call-option surface.
\vspace{0.5cm}


\end{center}

\vspace*{.1in} 

\section{Introduction}

One of the fundamental procedures in mathematical finance is the construction of the implied call surface (which is equivalent to modelling the implied volatility surface, by inversion of the Black-Scholes formula).
In practice, such a surface is only observed on a finite two-dimensional grid of maturity and strikes, i.e. at the points where market quotes are available. However, in order to price derivatives instruments involving values that fall outside the grid, it is necessary to be able to refine the resolution of the surface. 
As a first approximation, common practice to use spline interpolation between the points to recover a whole smooth surface. There are however some restrictions on the space of admissible solutions: call surfaces have to be subject to some no-arbitrage constraints across maturities and strikes (Carr-Madan\cite{CM05}, Davis-Hobson\cite{DH07}). In that regard, spline-interpolation of the call surface introduces arbitrage -- an undesirable feature.

Rather than focusing on call-surfaces, the literature has a tendency to deal with modelling the implied volatility surface directly.
As this problem is a very important one, it is not surprising many approaches have been proposed in the literature to address the modelling of the volatility surface. In Fengler \cite{F09}, an arbitrage-free smoothing of the call surface relying on least-square minimisation with smoothness penalty is suggested. See also Orosi \cite{O12} for a empirical evaluation of such a methodology. Another approach based on penalised $L^p$-norm was introduced in Fengler-Hin \cite{FH15}. In a recent article, Gatheral and Jacquier \cite{GJ14} proposed a parametrisation of the volatility surface and give conditions for the parametrised surface to be free of calendar-spread and butterfly-spread arbitrage. All these approaches allow to obtain an arbitrage-free surface, while aiming at being not too far from the original market quotes.

The inherent difficulty in these approaches is that the no-arbitrage conditions on the implied-volatility surface are highly nonlinear, while working with call surfaces gives --as we shall recall later-- only linear no-arbitrage constraints.

In this article, we propose a linear-programming based methodology that recovers a call surface matching exactly the market quotes up to user-specified tolerance and that is arbitrage-free. This approach is inspired by signal processing theory; the option call surface at a fixed time can be seen as a two-dimensional signal sampled on an initial grid (the marke points), and we will be aiming at recovering it on a finer grid while still perfectly matching the original observations and remaining arbitrage-free.

The classical result in sampling theory --the Nyquist-Shannon theorem-- states that to recover a signal exactly, it is sufficient to sample at twice the highest frequency contained in the signal. However, in certain cases such a rate is too much for one's purposes. In the signal processing community, compressed sensing has gained a lot of attention over the past years within several domains of application, notably MRI. A good introduction can be found in \cite{DDEK12}. The idea behind compressed sensing is that one does not need to have that many samples provided that the signal is \textit{sparse} in a given observation basis.
In this article, we inspire ourselves from the compressed sensing framework by writing the call-option surface problem as a constrained $l_1$-minimisation problem.

Section \ref{sec_theproblem} discusses the problem implementation. In particular, Subsection \ref{noarb_implementation} recalls the no-arbitrage conditions the solution has to satisfy, discusses what basis to take and the construction of the observation matrix. Subsection \ref{subsec_nonparam} recasts the problem in a nonparametric regression context, and subsection \ref{subsec_objfunc} treats of sparsity, oscillations in the signal and the objective function of the program. Finally, Subsection \ref{subsec_lp} give the final form of the optimisation problem and its recast as a simple linear-program.

An example of application is provided in Section \ref{sec_application}, where, starting from a limited number of call quotes around the money of the S{\&}P500, we recover a finer call grid and the corresponding implied-volatility surface, by inverting the Black-Scholes formula.

Applying the methodology to FX options on pegged currency (HKD/USD), we see that the methodology may fail to recover a sparse representation. Introducing structure-preserving functions in the observation matrix allows to circumvent this issue. This is the purpose of Section \ref{FX_application}.

\section{Problem setting and \texorpdfstring{$l_1$}{l1}-recovery}
\label{sec_theproblem}

Consider a set of maturities $\mathcal{T}$.
Given an underlying stock price $S$, a risk-free rate $r : T \mapsto r_T$ and a dividend function $q: T\mapsto q_T$, we denote the forward and the discount factor respectively as
\begin{equation*}
F(T) = Se^{(r_T-q_T)T},\quad D(T) = e^{-r_T T}.
\end{equation*}

In the rest of this chapter, in order to simplify the notations, we make the following assumption.
\begin{assumption}
\label{strike_assumption}
We assume below that for every maturity $T_i\in\mathcal{T}$, 
\begin{equation*}
K^{T_i}_j = K^{T_1}_j \frac{F(T_i)}{F(T_1)}.
\end{equation*}
 This also entails that there is the same number of strikes to recover on each time slice.
\end{assumption}

As a consequence, the $j$\textsuperscript{th} strike at maturity $T_i$ is written $K^{T_i}_j$. When the maturity is clear from context, we shall just write $K_j$ for $K^{T_i}_j$.

\subsection{call-surface recovery as a (constrained) nonparametric regression problem}
\label{subsec_nonparam}

From a practical point of view, perfectly matching the market mid-prices may be too much to ask for, or just impractical. Let us consider a few cases.
First, if there is initial arbitrage in one or more quotes, as sometimes happens (e.g. non-convexity in strike), not only is there no solution to the problem to begin with, as convexity will be lost on a smaller grid as well, but one \textit{does not} want to match the mid-price exactly. If one allows for a certain tolerance around the points, then a proper solution may be recovered.
Second, from a practical point of view, it may be preferrable to accommodate a framework that allows a user-specified tolerance. It can be the market bid-ask spread (as in liquid markets), or particular trader's needs/preferences.

The problem we aim at solving is now a little more precise. Taking the (vectorised) call-surface $C$, we want to find a function $f$ such that
$$
C(K,T) = f(K,T)+ \epsilon_1,
$$
where $f$ is an unknown function, and $\epsilon_1$ in an error term, which we seek to control and bound.

Now, while $f$ could be virtually any (continuous) function, we decompose it into a basis (or concatenation of bases) $(f_i)_{i=1}^\infty$,
$$
f = \sum\limits_{i=1}^{\infty} x_i f_i(T,K).
$$
\begin{remark}
The family $(f_i)_{i=1}^N$ is also sometimes called a \textit{dictionary}, or an aggregation of dictionaries, such as bases of polynomials, trigonometric functions, wavelets, etc.
\end{remark}

For obvious numerical reasons, it is necessary to truncate the sum:
$$
\tilde{f} = \sum\limits_{i=1}^{N} x_i f_i(T,K) + \epsilon_2,
$$
for some $N$, as well as to evaluate the $f_i$ on a grid. If we write $Q$ for the evaluation matrix of the $f_i$ on such a grid --- whose construction we shall detail below --- the initial relationship becomes
\begin{equation*}
C(K_i,T_j) = Q x + \epsilon_{1}+\epsilon_{2}.
\end{equation*}

We write $\epsilon :=\epsilon_1+\epsilon_2$ the general error term, which we wish to remain within pre-specified bounds.
Now is also the time to choose what decomposition basis to take. A natural family to be considered is polynomials in $T$ and $K$, more precisely, \textit{tensor orthonormal polynomials} in $T$ and $K$. Orthonormal bases are a privileged choice for their numerical stability. We define two (truncated) families of polynomials $(P_K^i)_{i=1}^{N_K}$ in $K$ and $T$ respectively.
\begin{enumerate}
\item The $P_T^i$ are orthonormal on the set of all maturities $(T_i)_{i=1}^{M_T}$:
\begin{equation*}
<P_T^i, P_T^j> = \int_{T_1}^{T_{M_T}} P_T^i(t) P_T^j(t) \mu_\mathcal{T}(\mathrm{d}T) = \delta_{ij},\qquad \mu_\mathcal{T}(\cdot) = \sum\limits_{i=1}^{M_T} \mathds{1}_{T_i}(\cdot).
\end{equation*}

\item Similarly, the $P_K^i$ are orthonormal on the grid of all considered strikes across maturities. Denoting by $\mathcal{K}$ the subset of strikes $K(T_i,j)_{i=1,\cdots,M_T}^{j=1,\cdots,M_K}$  where no strike appears more than once, we define:
\begin{equation*}
<P_K^i, P_K^j> = \int_{K_1}^{K_{M_K}} P_K^i(k) P_K^j(k) \mu_K(\mathrm{d}k) = \delta_{ij},\qquad \mu_K(\cdot) = \sum\limits_{\kappa_i\in \mathcal{K}} \mathds{1}_{\kappa_i}(\cdot).
\end{equation*}
\end{enumerate}

Let us now define the following matrices of $\mathbb{R}^{M_K\times N_K}$:
\begin{equation*}
Q_{mn} =
\begin{pmatrix}
P_T^n(T_m)P_K^1(K(T_m,1))&\cdots& P^n_T(T_m) P_K^{N_K}(K(T_m,1))\\
\vdots          &      & \vdots \\
P_T^n(T_m)P_K^1(K(T_m, M_K))&\cdots &P^n_T(T_m)P_K^{N_K}(K(T_m,M_K))\\
\end{pmatrix}.
\end{equation*}

Then the matrix of \textit{tensor polynomials} $Q\in \mathbb{R}^{(M_T \times M_K)\times (N_T\times N_K)}$ is simply defined by blocks as
\begin{equation*}
Q = \begin{pmatrix}
Q_{11}&\cdot &Q_{1\ N_T}\\
\vdots& &\vdots\\
Q_{M_T 1}& \cdots & Q_{M_T N_T}
\end{pmatrix}.
\end{equation*}

\subsection{Sparsity, \texorpdfstring{$l_1$}{l1}- and weighted-\texorpdfstring{$l_1$}{l1} recoveries}
\label{subsec_objfunc}
Now that we have determined what the basis and the observation matrix should be, we focus on the features $x$ should guarantee, namely \textit{sparsity}. We want the call-surface to be explained by as few parameters as possible. So we are seeking a parsimonious vector $x \in \mathbb{R}^N$, that is, a vector that has the smallest possible ``$l_0$-seminorm'':
$$
\|x\|_{0} := \sum\limits_{i=1}^N \mathds{1}_{x_i \neq 0}.
$$
Requiring parsimony of the solution is highly desirable, as it essentially requires that the solution does not overfit the data. We are aiming at obtaining a ``meaningful'' decomposition of $f$ in the basis $(f_i)_{i=1}^N$.

One problem remains though: problems of finding the sparsest solutions in the $l_0$ sense are in general \textit{NP-hard}. 

For example, NP-hardness of the $l_0$-recovery with quadratic constraints was proved by Natarajan \cite{N95}. See Elad \cite[p. 13-14]{E10} for a brief explanation of the issue with $l_0$-minimisation problems. Essentially, if a problem is NP-hard, then it means that there is currently no efficient algorithm (i.e. an algorithm with polynomial-time complexity) to solve it. 

Because of that NP-hardness, a way to address it is to consider the convex relaxation of the problem, and try to minimise the $l_1$ norm instead:
$$
\|x\|_{1} := \sum\limits_{i=1}^N |x_i|,
$$
i.e. minimise the total energy of the signal. This is referred to as \textit{$l_1$-recovery}.

\begin{remark}
\label{compressed_sensing_refs}
We point out that conditions guaranteeing equivalence of $l_0$ and $l_1$ solutions is the \textit{raison d'\^etre} of the research area known as \textit{Compressed Sensing}. There is a vast literature that shows that in the case of a random basis satisfying the so-called Restricted Isometry Property, it is possible to find equivalence of these two solutions. A rigorous description is beyond our scope here; we refer the interested reader to Davenport et al. \cite{DDEK12} for an introduction, as well as a (highly non-exhaustive) list of articles of interest: Cand\`es-Donoho-Sanders \cite{CDS98}, Cand\`es \cite{C08},Cand\`es-Tao \cite{CT05}, Donoho-Elad \cite{DE03}. 
\end{remark}

\begin{remark}
Choosing an $l_1$-norm means that we would rather recover a solution that does not allow for as many non-zero components with small amplitude as, for example, the $l_2$-norms. However, note that, at that stage, this does not prevent us from obtaining a solution involving high-degree -- hence highly oscillating -- polynomials. The appearance of such high-degree polynomials can be seen as overfitting; taking financial consequences into account, this would entail obtaining noisy surfaces when looking at the state-price distribution of the underlying, or computing the local-volatility surface.
\end{remark}

It turns out that weighting the $l_1$ norm gives a way of enhancing sparsity and reducing oscillations, as mentioned in Rauhut-Ward \cite{RA15}; i.e. replacing the $l_1$-norm by a norm of the form
$$
\|x\|_{Wl_1} = \sum\limits_{i=1}^N w_i \cdot x_i,
$$
where the $w_i$ are predetermined weights. Rauhut-Ward consider various weights, and setting $w_i = i$, i.e. penalising polynomials of high order, provides a substantial improvement in reducing the oscillations in their dataset; this can be seen as a smoothing penalty in the solution in the context of $l_1$-recovery. Based on their results, we shall also considers weighted objective norms of the form
$$
\|x\|_{obj} = \sum\limits_{i=1}^N w_i \cdot x_i.
$$

\subsection{No-arbitrage constraints on the solution}
\label{noarb_implementation}

We recall here the conditions for a call-option surface to be arbitrage free and see how these constraints implement in our problem.
For pricing and hedging purposes, obtaining the arbitrage-free implied-volatility surface (or equivalently the call-option surface) is a problem of practical importance. While no-arbitrage conditions on the implied volatility surface are highly nonlinear (Roper~\cite{R10}), absence of arbitrage on the call-option surface is guaranteed by the following set of conditions (see e.g. Fengler~\cite{F09}): we denote by $F(T)$ the forward value of the underlying at time $T$ and by $D(T)$ the value of a unit zero-coupon bond expiring at maturity $T$.

\begin{enumerate}
\label{cond_noarb}
\item Convexity: for all $T$, the mapping $K\mapsto C(K,T)$ is convex. The reason is the following: Consider three call options $C(K_1)$, $C(K_2)$ and~$C(K_3)$ such that $K_1<K_2<K3$. One can then construct the \textit{butterfly} derivative instrument
\begin{equation}
\label{asymmetric_butterfly}
B_f(K_1,K_2,K_3) := (K_3-K_2)C(K_1)-(K_3-K_1)C(K_2)+(K2-K_1)C(K_3),
\end{equation}
whose payoff is given in Figure \ref{asymmetric_butterfly_fig}.
\begin{figure}[H]
\begin{center} 
\includegraphics[width = 5cm, height = 5cm]{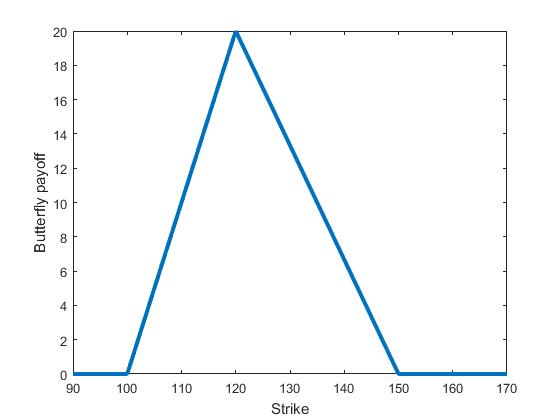}
\caption[The payoff of an asymmetric butterfly.]{The payoff of an asymmetric butterfly, with $K_1 = 100$,$K_2=120$ and $K_3=150$.}
\label{asymmetric_butterfly_fig}
\end{center}
\end{figure}
As one can see, this payoff is nonnegative. A breach in convexity entails that one can find three strikes and create such a butterfly position for a nonpositive price, hence an arbitrage.
\item Absence of Calendar-spread arbitrage: for two maturities $T_1<T_2$ and any strike $K^{T_1}$, denote $K^{T_2}$ its forward value at maturity $T$, namely
\begin{equation*}
K^{T_2} = K^{T_1} \cdot \frac{F(T_2)}{F(T_1)}.
\end{equation*} 
Then we need the mapping $T\mapsto C(K^T,T)$ to be non-decreasing.
To see that, take two strikes $C(K^{T_1},T_1)$ and $C(K^{T_2},T_2)$ with $T_1< T_2$ and 
\begin{equation*}
K^{T_2} = K^{T_1}F(T_2)/F(T_1).
\end{equation*}
Then, if one were to buy a call $C(K^{T_2},T_2)$ and sell an amount
\begin{equation*}
D(T_2)F(T_2)/(D(T_1)F(T_1))
\end{equation*} 
of calls $C(K_1,T_1)$.
At time $T_1$, if the underlying value $S(T_1)$ is less than $K^{T_1}$, then the final payoff is just $C(K^{T_2},T_2)>0$; otherwise, the position at $T_1$ is
\begin{equation*}
C(K^{T_2},T_2)-\frac{D(T_2)F(T_2)}{D(T_1)F(T_1)}(S(T_1)-\frac{F(T_1)}{F(T_2)}K^{T_2}),
\end{equation*}
which is equal to a put option with maturity $T_2$ and strike $K^{T_2}$ by put-call parity. A put option having a nonnegative payoff, this position must also be nonnegative. So in either case, we need this position to be nonnegative.
\item Monotonicity: we need
\begin{equation*}
-1/(D(T)F(T))\leq \frac{\partial C}{\partial K}(K,T) \leq 0.
\end{equation*}
The left-hand side inequality means that an infinitesimal increase in the strike --- which induces the same infinitesimal decrease in the final payoff --- must produce at most this (discounted) decrease on the current price of the call. The right-hand side inequality means that, all else equal, a call with higher strike will always have a lower payoff than a call with lower strike.
\item Payoff bounds: for $S$ the current price of the underlying, for all $K,T$, we need
\begin{equation*}
(D(T)F(T)S-KD(T) )^+\leq C(K,T)\leq SD(T)F(T).
\end{equation*} 
The first inequality means that the value of the call today must be nonnegative (since its payoff will always be nonnegative) and higher than the value of a forward contract with the same strike. The second inequality states that the value of a call with always yield less than simply holding the stock, adjusted by dividends. Indeed, currently having
\item Final value: $C(K,0)= (S-K)^+$, i.e. ``the value of the is the value of its payoff''.
\end{enumerate}

\subsubsection{Absence of butterfly-spread arbitrage}

We first consider the constraints that will guarantee that assumption 1 is satisfied, namely that the recovered finer surface is convex.
In particular, at the extremal points of the call-surface $K_1$ and $K_{M_K}$, the no-butterfly/convexity requirements become the following, where $S$ denotes the price of the underlying stock:
\begin{align*}
&SD(T)F(T)-\frac{K_2}{K_2-K_1} C(K_1,T)  + \frac{K_1}{K_2-K_1} C(K_2,T)\geq 0\\
&C(K_{M_K-1},T) - C(K_{M_K},T) \geq 0.  
\end{align*}

For $K_1$, this corresponds to writing the asymmetric butterfly condition where one call has strike zero (and so its value is just the underlying stock, adjusted by dividends; equivalently the discounted forward). For $K_n$, it corresponds to writing the asymmetric butterfly condition where one call has infinite strike, and so has value zero.

As has been seen before, the (vectorised) recovered call-surface $C$ is written 
\begin{equation*}
C = Qx.
\end{equation*} 
Then, the butterfly condition becomes
$$
B_f Q x \geq R,
$$
where
\begin{equation}
B_f = \begin{pmatrix}
B_{f,1}& & & &0\\
&B_{f,2}& & &\\
& & \ddots& &\\
0& & &      & B_{f,M_T}
\end{pmatrix},
\end{equation}
with
$$
B_{f,i} :=\begin{pmatrix}
 -\frac{K_2}{K_2-K_1}  &\frac{K_1}{K_2-K_1}    & 0     &       &       &\\
1&-\frac{K_3-K_1}{K_3-K_2}&\frac{K_2-K_1}{K_3-K_2}& 0     &\cdots &0\\
0      &1&-\frac{K_4-K_2}{K_4-K_3}&\frac{K_3-K_2}{K_4-K_3}&  0    &\cdots\\
0      & 0     &1&-\frac{K_5-K_3}{K_5-K_4}&\frac{K_4-K_3}{K_5-K_4}& \\
0      & 0     & 0     &\ddots&\ddots&\ddots&\\
0      &      & \cdots     &0&-1&1   
\end{pmatrix}. 
$$
and with 
$$
R_i=\begin{cases}
-SD(T)F(T) \text{ if } i \equiv 1 \text{ mod } M_K \\
0 \text{ otherwise}.
\end{cases}
$$

\subsubsection{Absence of calendar-spread arbitrage}

Concerning the absence of calendar-spread arbitrage, the condition reads as follows. Denoting the discount factor up to time $t$ by $D(t)$ and $F(t)$ the forward price up to time $t$, we require, for all $K_i$,
$$
C(K^{T_{j+1}}_i,T_{j+1})-\frac{D(T_{j+1})F(T_{j+1})}{D(T_j)F(T_{j})}C(K^{T_j}_i,T_{j})\geq 0.
$$
Consider the matrix $G\in\mathbb{R}^{(M_K\times M_T-1)\times(M_K\times M_T)}$. For indices $i\in\{1,\cdots,M_K\}$,
$j\in \{1,\cdots,M_T-1\}$ and $l\in\{1,\cdots,M_K\}$,

$$G((j-1)\times M_K+i,l) = \frac{D(T_{j+1})F(T_{j+1})}{D(T_j)F(T_{j})}\delta_{\{(j-1)\times M_K+i=l\}} -\delta_{\{j\times M_K+i=l\}}.$$
Then, the no arbitrage condition becomes
$$
G Q x  \leq 0.
$$

\subsubsection{Absence of vertical-spread arbitrage}
Regarding the constraints
$-D(T)\leq \frac{\partial C}{\partial K} \leq 0$, these can be rewritten as two matrix inequalities.
Write $H\in \mathbb{R}^{(M_K\times M_T)^2}$ with
$$H(i,j) =  \mathds{1}_\{i=j\} -\mathds{1}_\{i-1=j\},$$
and
$$
U_b(i) = D(T_j)F(T_{j}) S \mathds{1}_{\{(i-1 \equiv 0 \text{ mod } M_K)\}}.
$$
Then the absence of vertical spread rewrites as
\begin{align*}
H Q x \leq U_b.
\end{align*}

\subsubsection{Constraints on the bounds}
It can be checked (Fengler \cite{F09}) that if all other arbitrage conditions (butterfly, vertical spread and calendar spread) are already enforced, and under the assumption \ref{strike_assumption}, then the remaining restriction
$$
(F(T_i)-K^{T_i}_j)^+\leq C(K^{T_i}_j, T_i)\leq F(T_i)
$$
can simply be enforced by imposing a dimension-one constraint on the lowest point of the surface:
$$
C(K^{T_1}_{M_K},T_1) \geq (F(T_1)-K^{T_1}_{M_K})^+.
$$

As a consequence, when working on the call-surface, all no-arbitrage conditions are linear and can all be merged into one no-arbitrage matrix inequality
\begin{equation}
\label{noarb_mat_form}
L x  \leq J,
\end{equation}
with $L \in\mathbb{R}^{(M_K\times(4M_T-1)+1) \times (N_K\times N_T)}$, and $J\in \mathbb{R}^{M_K\times(4M_T-1)+1}$.

We now have the elements to state the $l_1$-recovery problem.

\subsection{Weighted \texorpdfstring{$l1$}{l1}-recovery problem and LP equivalence}
\label{subsec_lp}

We denote by $C^o$ the vector of original market mid-prices quotes. The only thing left to discussing is the form taken by the weighted $l_1$ norm appearing in the objective function of our minimisation problem. The purpose of weighting the objective function is so that the optimiser will favour solutions 

Recall that by order of a polynomial, we refer to its degree plus one.
A tensor polynomial should be penalised a function of its degree/order: the higher its degree/order, the more penalised it should be, and so the higher the $w_i$ must be.
For example, the constant polynomial has order one in strike and in maturity, hence its corresponding weight will be $2$. The polynomial $K^2T$ has order three in strike and order $2$ in maturity, and so its corresponding weight will be $5$. 
Write $x\in\mathbb{R}^{N_K\times N_T}$ for the signal vector we aim at recovering and consider an element $x_i$ thereof, with $i=(y-1)\cdot N_K+z$, $y\in \{1,\cdots,N_T\}$ and $z\in \{1,\cdots,N_K\}$. Then, the weight associated with $x_i$ will be $w_i=y+z$.

Based on the previous sections, the problem we are solving can be rewritten as:
\begin{equation}
\label{wl1}
\tag{WL1}
\mathrm{(W-L_{1})}:=
\begin{cases} 
\min\limits_{x \in \mathbb{R}^{N_K\times N_T}} \sum\limits_{i=1}^N w_i\cdot x_i,\\
\text{such that}\\
|A x - C^o|\leq \epsilon,\\
L x \leq J.
\end{cases}
\end{equation}

There exists an easy method to solve this problem: notice it can be recast as a linear program by making the change of variable $x = u-v$, with $u,v \geq 0$. In standard form, this becomes
$$
\mathrm{(LP)}:=
\begin{cases} 
\label{linprog}
\min\limits_{u,v \in \mathbb{R}^{N_K\times N_T}} \sum\limits_{i=1}^N w_i (u_i+v_i),\\
\text{such that}\\
|A (u-v)-C^o|\leq \epsilon,\\
 L (u-v) \leq J,\\
u,v \geq 0,
\end{cases}
$$
meaning that the recovery problem can be solved using classical linear-programming optimisers. We refer to Luenberg-Ye \cite{LY08} for background on linear programs as well as related numerical algorithms.

\section{Application to the S\& P500 call-surface}
\label{sec_application}
\begin{figure}[htbp]
\begin{center} 
\includegraphics[width = 7cm, height = 6cm]{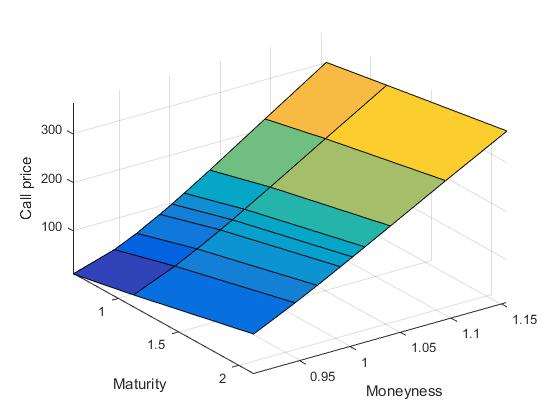}
\caption[Plot of the original S{\&}P500 call prices on 30 Oct. 2015 from CBOE.]{Original data, consisting in S{\&}P500 call prices at 9 strikes and 3 maturities on 30-Oct-2015. Data obtained from the CBOE website.}
\label{orig_data}
\end{center}
\end{figure}

Let us apply the above procedure to a small sample of mid-price call options of the S{\&}P500 priced on Oct 30. 2015, displayed in figure \ref{orig_data}. 
Before applying our methodology, we illustrate the caveats cubic spline interpolation may produce in the next Subsection.
\subsection{Fitting splines}
\begin{figure}[H]
\begin{minipage}{.5\linewidth}
\centering
\includegraphics[width = 7cm, height = 6cm]{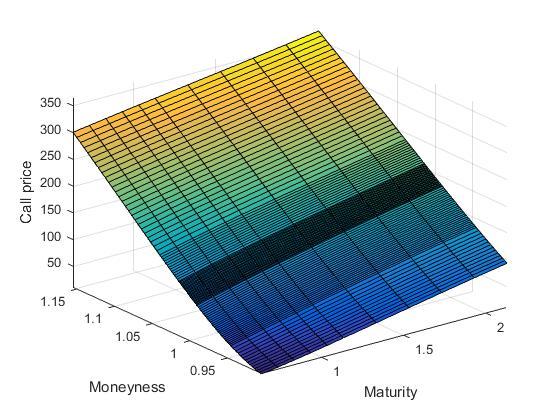}
\end{minipage}%
\begin{minipage}{.5\linewidth}
\centering
\includegraphics[width = 7cm, height = 6cm]{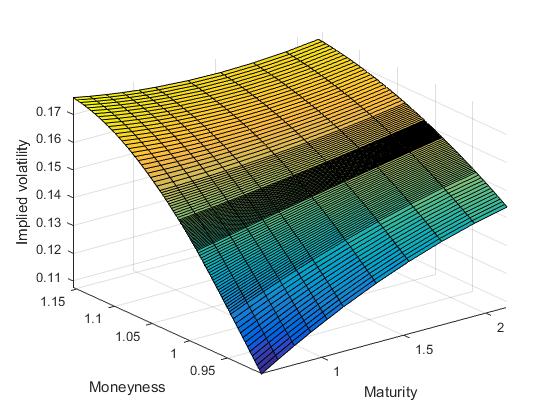}
\end{minipage}\par\medskip
\centering
\includegraphics[width = 7cm, height = 6cm]{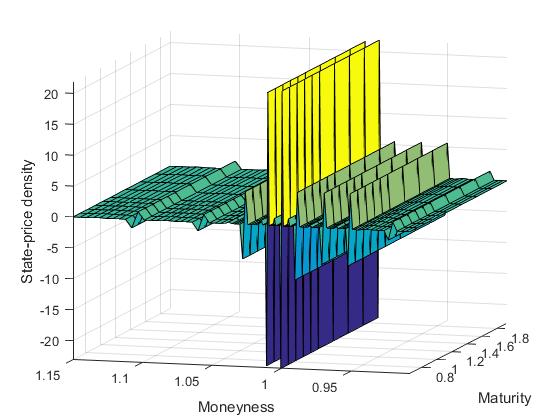}
\caption[Recovered S{\&}P500 call and implied-volatility surfaces using cubic-splines]{Top-left: call-surface recovered from spline interpolation on implied volatilities. Top-right: Implied-volatility surface recovered from spline interpolation. Bottom: associated state-price density. The density takes negative values, indicating towards butterfly arbitrage. Each slice of the interpolated surface has $M_K=104$ strikes, for a total of $M_T=11$ slices.}
\label{call_surface_match_spline}
\end{figure}

Before proceeding with our methodology, we shall attempt cubic-spline interpolation. Cubic splines are perhaps the most straightforward way to interpolate a volatility surface, but they are known for oscillating and introducing arbitrage. While Figure \ref{call_surface_match_spline} seems to yield a fairly good recovery, the associated state-price density oscillates and takes negative values, exhibiting butterfly arbitrage.

\subsection{Call-surface recovery}
\label{sec_recov}
Using the same dataset, we now proceed to using our $l_1$-recovery procedure. The code has been written in MATLAB, and we used the GUROBI \cite{gurobi} optimiser.

We point out that orthonormal polynomials behave badly on the boundaries of the interval of interest. The way we deal with it is by expanding the domain of maturities and strikes on which to perform the recovery, thus controlling possible explosions outside our original domain.
No-arbitrage constraints are also enforced on the extended domain, so as to allow for extrapolation in the vicinity of the original domain; this last point, however, can be easily relaxed.

\begin{figure}[H]
\begin{center} 
\begin{align*}
\includegraphics[width = 7cm, height = 6cm]{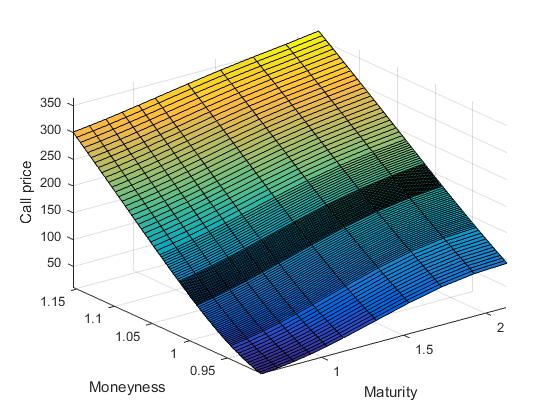}
&\includegraphics[width = 7cm, height = 6cm]{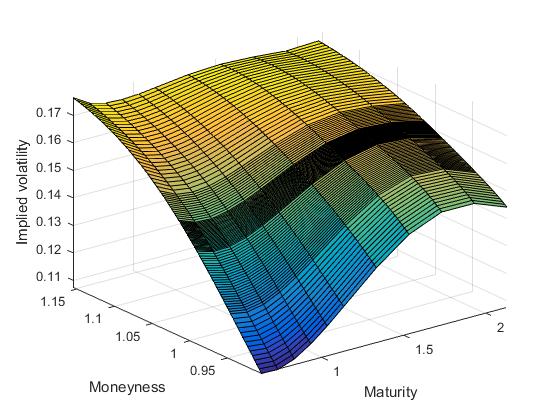}
\end{align*}
\caption[Recovered S{\&}P500 call and implied-volatility surfaces using $l_1$-recovery]{Left: call-surface recovered from weighted-$l_1$ recovery. Right: corresponding implied volatility surface. relative tolerance $\epsilon \equiv 5\times 10^{-4}$, $M_K=104$, $N_K=14$, $M_T=11$,$N_T=7$.}
\label{call_surface_match}
\end{center}
\end{figure}

In order to recover the call prices on the finer grid $(K_i,T_j)_{i=1,\cdots,M_K,j=1,\cdots,M_T}$, applying the algorithm returns the following call-surface, as well as the associated implied-volatility surface (Figure \ref{call_surface_match}). We recover slices for $11$ maturities, each consisting of $104$ strikes. The basis has been truncated to allow tensor polynomials to have up to degree $13$ in strike (corresponding to polynomial order in strike~$N_K=14$) and degree $6$ in time (corresponding to polynomial order in strike ~$N_T=7$).

In particular, we check that we indeed lie withing the pre-specified bounds:
\begin{figure}[H]
\centering 
\includegraphics[scale=0.7]{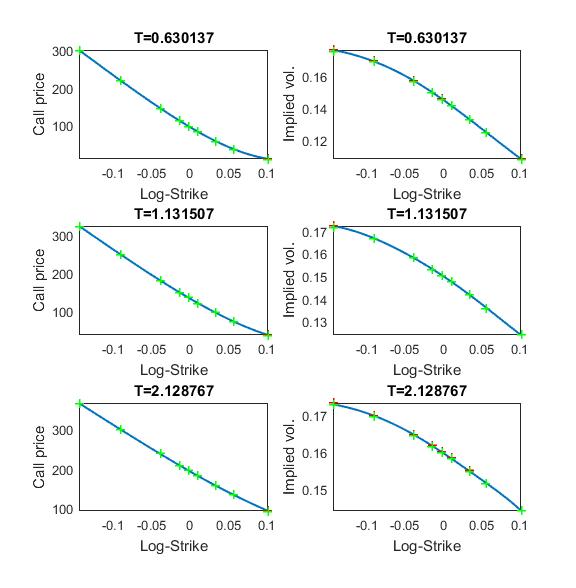}
\caption[Slice plot of the recovered call prices and implied volatilities]{Plots of the fitted Call prices (left graphs) and corresponding implied volatilities (right graphs) at market maturities. Green and red markers are very close and indicate the lower and upper tolerance at market points respectively.}
\label{l1_slices}
\end{figure}

\begin{figure}[H]
\begin{center} 
\begin{align*}
\includegraphics[width = 7cm, height = 6cm]{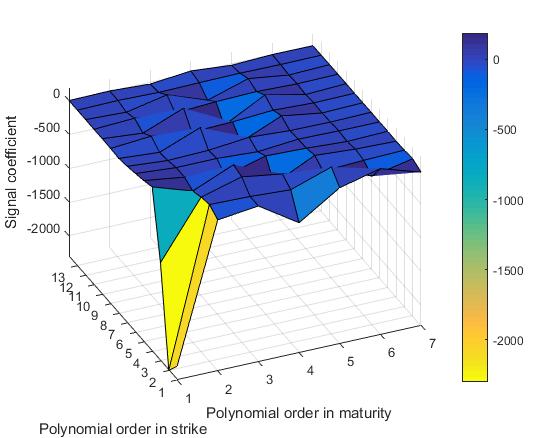}&\includegraphics[width = 7cm, height = 6cm]{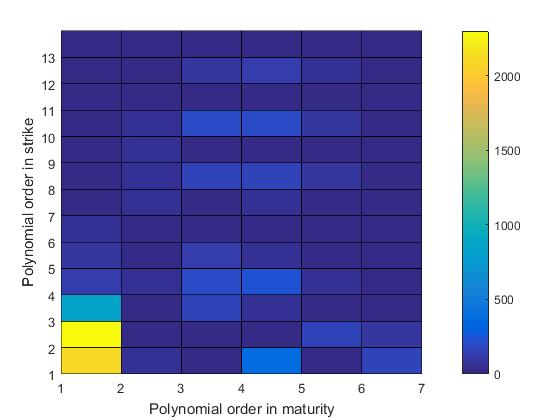}
\end{align*}
\caption[Signal coefficients obtained from the $l_1$-recovery procedure on the S{\&}P500 data.]{Plot of the coefficients $x$ obtained from the $l_1$-recovery procedure on the S{\&}P500 data. On the right: the absolute values of the signal, viewed from above.}
\label{signal}
\end{center}
\end{figure}

We also take a look at the sparsity of the underlying signal. For that purpose, we plot the signal $x$ in the form of an $N_K\times N_T$ matrix in Figure \ref{signal}.

As can be seen, the signal retains some sparsity features, where the highest coefficients are concentrated around polynomial indices with small degree in $K$ and $T$.

Finally, we take a look at the state-price density and local volatility surfaces. In our context, these quantities have a purpose for evaluating overfitting, by inspecting the shapes thereof.

\begin{figure}[H]
\begin{minipage}{.5\linewidth}
\centering
\includegraphics[width = 7cm, height = 6cm]{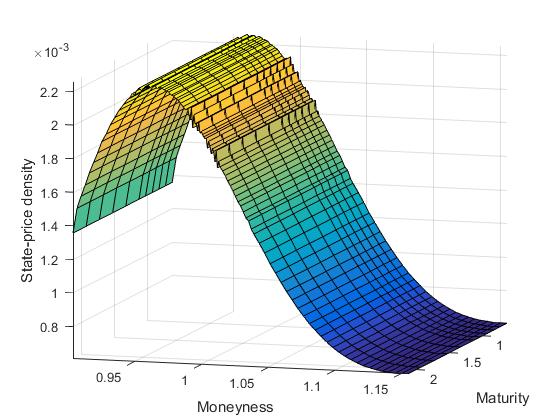}
\end{minipage}%
\begin{minipage}{.5\linewidth}
\centering
\includegraphics[width = 7cm, height = 6cm]{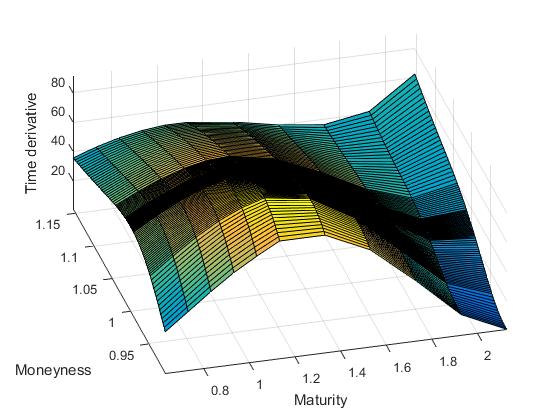}
\end{minipage}\par\medskip
\centering
\includegraphics[width = 7cm, height = 6cm]{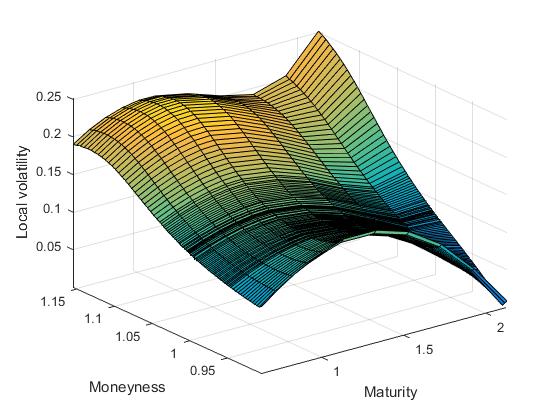}
\caption[Density, time-derivative and local-volatility surfaces recovered from the $l_1$-recovery procedure on the S{\&}P500 data.]{Top-left: surface plot of the state-price density, or equivalently $\partial^2 C/\partial K^2$. Top-right: surface plot of the first time derivative $\partial C/\partial T$. Bottom: the corresponding recovered local-volatility surface.}
\label{local_vol}
\end{figure}

The local-volatility surface, in the case of neither interest rate nor dividends, is given by
\begin{equation*}
\sigma_{loc}(K,T) = \frac{1}{K}\sqrt{2\frac{\partial^2 C}{\partial K^2}\Big/ \frac{\partial C}{\partial T}},
\end{equation*}
is given in Figure \ref{local_vol}.
Figure \ref{local_vol} plots the quantities $\partial C/\partial T$, $\partial^2 C/\partial K^2$ and the corresponding local volatility $\sigma_{loc}$.  Overfitting would typically result in a noisy local-volatility surface. The fact we obtain a bell-shaped curve for the state-price density, as has been documented in the literature (e.g. in \cite{FH15}), suggests that the fit is reasonable.

\begin{remark}
The number of basis functions plays an important role in the surface recovery. Of course, if one truncates the basis too much, we may fail to recover a solution. If one does not truncate enough, one should be fine in theory, as adding functions would result in a solution at least as good. However, we have noted that expanding the space too much may lead GUROBI to a suboptimal solution, i.e. a solution whose objective function value is actually higher than one obtained with less basis functions available. 
\end{remark}

\section{An FX application: pegged currencies}
\label{FX_application}
As described in Clark \cite{C11}, implied-volatility market quotes are usually given for straddles and butterflies and quoted in deltas, rather than for vanilla calls and puts and quoted in strikes. Conversion is directly available through Bloomberg; so our starting point shall be the bid/ask implied volatilities of vanilla calls given as a function of the maturity and strike (rather than the Delta).

Our currency pair of interest will be the HKD/USD, i.e. the value of Hong Kong Dollar in U.S. Dollars. The Hong Kong dollar has the specificity of being \textit{pegged} to the US Dollar, meaning that it is only allowed to move inside a fixed range from the U.S. Dollar. At the time of the writing, the Hong Kong dollar is such that one U.S. dollar has to be worth at least $7.75$ HKD and at most $7.85$ HKD.

The dataset we shall consider for the rest of this section is the set of bid-ask vanilla implied volatilities of January 19 2016, with maturities ranging from one week to two years, and five strikes per maturity, obtained from Bloomberg. From the same source, we also obtain the Hong Kong and United States interest rates, so that we can convert the implied volatilities into option prices.

In this section we shall aim at calibrating this dataset with both a parametric approach from the literature and the nonparametric approach described in the previous section.

For a parametric benchmark, Surface Stochastic Volatility Inspired parametrisations (SSVI) models present an attractive framework, and are reported to perform well on equity such at the S\&P500; see Gatheral-Jacquier \cite{GJ14}. 
As a benchmark, we apply their SSVI square-root parametrisation, which guarantees an arbitrage-free surface up to very long, impractical expiries. 
 While one can see that the fit works very well overall, one can observe that at higher maturities, the SVI fit lies outside the bid-ask spread at the 25-Delta Put quote.

\begin{figure}[H]
\centering
\includegraphics[width = 14cm, height = 14cm]{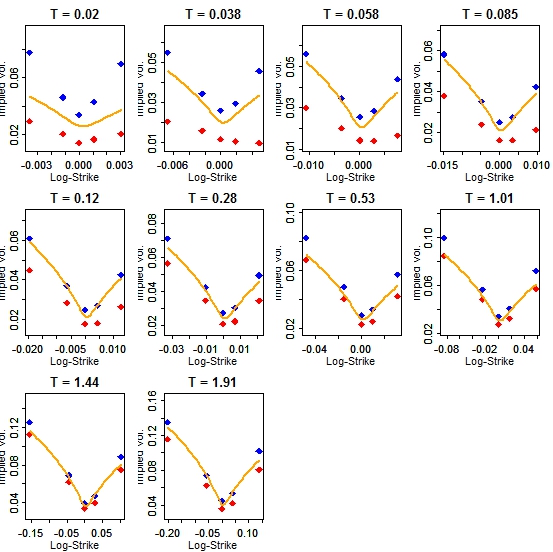}
\caption[Fitting of the Square-root SSVI model to HKD/USD mid-implied-volatilities (19/01/2016, source: Bloomberg) for maturities ranging from one week to two years, using Gatheral's R-script.]{Fitting of the Square-root SSVI model to the HKD/USD mid implied-volatilities for maturities ranging from one week to two years, using Gatheral's R-script (Data source: Bloomberg). On each picture, the blue curve is the fitted SVI smile. Red diamonds indicate the bid market data and blue diamonds indicate the ask marked data. For the last two maturities and the 25-Delta Put strike, the calibrated SVI smile lies outside of the bid-ask spread.}
\label{SVI_calibration}
\end{figure}

Regarding these observations, and in the case where one would need to find a solution guaranteed to lie within the bid-ask spread for any market-quote, there may be a case for applying our nonparametric methodology presented in the last chapter.

For more than eight maturities, our previous algorithm fails to converge.

\begin{figure}[H]
\begin{center} 
\begin{align*}
\includegraphics[width = 7cm, height = 6cm]{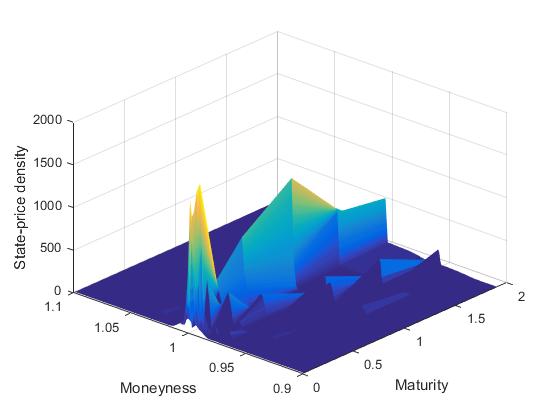}
&\includegraphics[width = 7cm, height = 6cm]{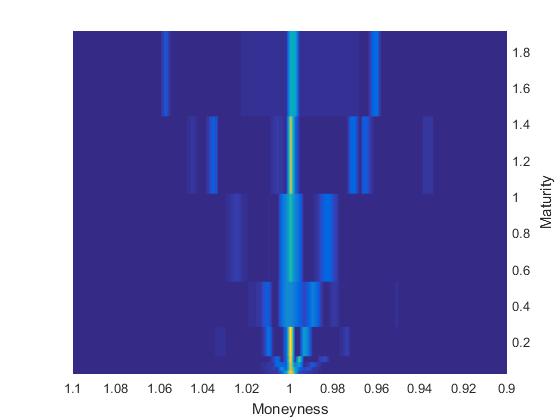}
\end{align*}
\caption[Density obtained from using the tensor-polynomials $l_1$-recovery procedure on HKD/USD bid-ask call prices (19. Jan 2016) for maturities ranging from one week to two years.]{On the left, the recovered state-price density using a basis of tensor polynomials. On the right: the same picture, viewed from above: the corresponding density is extremely noisy, with spikes across the whole surface.}
\label{distrib_tensor}
\end{center}
\end{figure}

Even with only eight maturities, when inspecting the state-price density (Figure \ref{distrib_tensor}),  it is extremely noisy. This means that the optimiser has to use high-degree polynomials in order to constrain the shape of the final solution, resulting in a very noisy state-price density surface. In particular, it gives probability mass to strikes which lie outside the pegging . Essentially, the solution can be seen as emulating a piecewise linear fitting. Another way to state it is that we failed to recover a \textit{sparse} solution.

This can be interpreted in two ways. The main one is that even with a limited amount of maturities, it turns out that, on this particular dataset, polynomials may just not be a basis in which the call option surface is sparse. Visually inspecting the density plot, one can see that the surface is fairly steep: from a linear part in the money, it quickly breaks into an almost-flat part when strikes increases to out-of-the-money: a high number of polynomials would be required to fit the data.

The second may be because of the difficulties the optimiser may encounter while optimising simultaneously in strike and maturity. Since time interpolation across strike-levels makes little sense in the FX market, it may be sensible to focus on strike optimisation.

Following these observations, we shall relax the problem in both ways, by allowing all slices to have different projections, and also find a basis which satisfies our sparsity requirements.

\subsection{Problem reformulation}
\label{prob_reform}
While polynomials allow one to fit any continuous function (by the Weierstrass theorem, see e.g. \cite[Theorem 7.26.]{R76}), a high number of them may be required in order to provide a good approximation.

However, in our problem formulation, and unlike least-squares procedures acting on the call-prices, we have the liberty of choosing beforehand the projection basis. In particular, we can construct functions that would naturally be close to the original bid and ask data points, while satisfying shape constraints such as convexity.

The \textit{modus operandi} is the following. For every slice, we construct an ad-hoc set of functions in the following way.

First, we take a linear interpolation (and possible extrapolation) of the bid prices onto an equidistant grid of strikes, where the bounds are the minimal and maximal value of the grid we want to recover the call prices on at the end. A priori, and if there is no butterfly arbitrage, this function is convex. If there is arbitrage, this function is at least locally convex and can be adjusted.

We would like to be able to preserve convexity as much as possible, while having smooth basis functions. In order to do so, we introduce the following \textit{mollifier}.

\begin{definition}
A \textit{mollifier} $\phi$ on $\mathbb{R}^n$, $n\geq 1$ is an infinitely differentiable function from $\mathbb{R}^n$ to $\mathbb{R}$ such that
\begin{enumerate}
\item $\phi$ has compact support,
\item \begin{equation*}
\int_{\mathbb{R}^n} \phi(x) \mathrm{d}x = 1,
\end{equation*}
\item In the sense of distributions,
\begin{equation*}
\lim_{\alpha \to 0} \phi^{\alpha}(x) := \lim_{\alpha \to 0} \alpha^{-n}\phi(\frac{x}{\alpha}) = \delta(x),
\end{equation*}
where $\delta$ denotes the Dirac function at zero:
\begin{equation*}
\delta(x) = \begin{cases}
+\infty \text{ if } x= 0,\\
0 \text{ otherwise.}
\end{cases}
\end{equation*}
\end{enumerate}
\end{definition}

\begin{example}
A standard mollifier from $\mathbb{R}$ to $\mathbb{R}$ is given by:
\begin{equation*}
\phi_s(x) = \begin{cases}
K \exp\left(\frac{1}{|x|^2-1}\right)\text{ if } x < 1,\\
0 \text{ if } x  \geq 1.
\end{cases}
\end{equation*}
\end{example}
where $K$ is a constant such that $\phi_s$ integrates to one over $\mathbb{R}$.

\begin{figure}[H]
\begin{center}
\includegraphics[width = 7cm, height = 6cm]{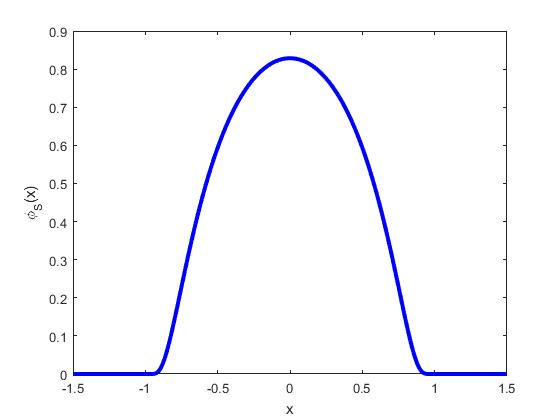}
\caption[The standard mollifier]{The standard mollifier in one dimension, $\phi_s:x \mapsto \exp(1/(x^2-1))\mathds{1}_{|x|<1}$}
\label{standard_mollifier}
\end{center}
\end{figure}

Then, in the following, we shall refer to the \textit{mollification} of a function as follows.
\begin{definition}[Mollification of a function]
Let $f:\mathbb{R}\to \mathbb{R}$ be a continuous function. We call $\widetilde{f}^\alpha$ the \textit{mollification} of $f$ with parameter $\alpha$ the function 
\begin{equation}
\label{mollification}
\widetilde{f}^\alpha(x) = \int_{\mathbb{R}} f(y)\phi^{\alpha}_s(x-y) \mathrm{d}y.
\end{equation}
\end{definition}

Since $\phi^{\alpha}\in C^{\infty}(\mathbb{R};\mathbb{R})$, it is easy to see that $\widetilde{f}^\alpha$ is $C^\infty(\mathbb{R};\mathbb{R})$ (see e.g. Hirsch \cite{H97} for standard properties of mollifiers). We also refer to Ghomi \cite{G02} for the application of mollification for convex-function smoothing.

\begin{definition}[Structure-preserving function]
\label{struct_pres_func}
By \textit{structure-preserving function}, we refer to any $C^2(\mathbb{R};\mathbb{R})$-function that takes as parameters the bid-ask original market data. For example, the mollification of the mid-price is such a function.
\end{definition}

\begin{figure}[H]
\begin{center}
\includegraphics[width = 7cm, height = 6cm]{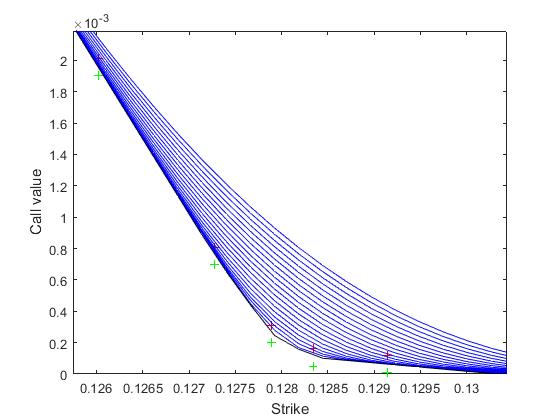}
\caption[The mollification of the HKD/USD call mid-price curve for various smoothing parameter values.]{Plot of the mollified mid-price for the maturity three weeks and various values of $\alpha$. As $\alpha$ increases, the mollified mid-price curve increases. The lowest curve (in black) represents the un-mollified original market mid curve. Green and red crosses indicate the market bid and ask prices respectively.}
\label{mid-mollification}
\end{center}
\end{figure}

Figure \ref{mid-mollification} displays mollifications of the mid-price linear interpolation for different values of $\alpha$. One can indeed see that for $\alpha$ small enough, $\widetilde{f}^\alpha$ converges to $f$ (a consequence from $\phi^\alpha$ converging to the Dirac function in the sense of distributions when $\alpha\to 0$). Essentially, the mollification is a weighted moving-average procedure. In particular, it can be seen on this particular example that for $\alpha$ small enough, the mollification lies within the bounds of the bid-ask spread.  One would reach the same conclusions starting for example from the mid-price linear interpolation instead of the bid.

We now have all the elements to restate the amendments to the problem from the previous chapter; we ``disentangle'' the surface and allow each slice to have its own basis decomposition (instead of having one tensor basis decomposition for the whole surface).

So, the problem rewrites as follows.

Letting $\phantom{m}^t z$ denote the transpose of a vector $z$, we aim at finding
\begin{equation*}
x :=\phantom{m}^t(x_{1}^{T_1}, x_{2}^{T_1}, \cdots, x_{N_K}^{T_1}|x_{1}^{T_2},\cdots,x_{N_K}^{T_2}|\cdots|x_{1}^{T_{M_T}},\cdots, x_{N_K}^{T_{M_T}})\in \mathbb{R}^{N_K\times M_T}, 
\end{equation*}
that solves
\begin{equation*}
\mathrm{(L{1}_{FX})}:=
\begin{cases} 
\min_{x \in \mathbb{R}^{N_K\times M_T} } \sum\limits_{i=1}^{N_K\times M_T} x_i,\\
\text{such that}\\
|A x - C^o|\leq \epsilon,\\
L x \leq P,
\end{cases}
\end{equation*}
with $A$, $L$ and $P$ to be defined below. First, we introduce the basis matrix~$Q$, defined by blocks:
\begin{equation*}
Q = \begin{pmatrix}
Q_1& & & &0\\
&Q_2& & &\\
& & \ddots& &\\
0& & &      & Q_{M_T}
\end{pmatrix}\in\mathbb{R}^{(M_T\times M_K)\times(M_T\times N_K)},
\end{equation*}
with all $Q_i\in \mathbb{R}^{M_K\times N_K}$. Every column of $Q_i$ corresponds to the evaluation of one function at the various pairs of strikes and maturities present in our grids. In our context, it can be either the structure-preserving function described in Definition \ref{struct_pres_func} evaluated at every strike present on slice $i$ for a certain mollification parameter $\alpha$, or a low-order polynomial.

As a consequence, the $(M_K\times i +j)$\textsuperscript{th} row of $Q$, when multiplied by $x$, will return the value of the call with maturity $T_i$ and strike $K_j^{T_i}$.

 Then $A$ is defined as the submatrix of $Q$ consisting of the rows where the pair $(K,T)$ corresponds to a market-quoted call option.

$L$ is the matrix containing all no arbitrage conditions, defined in a similar way as \ref{noarb_mat_form}. The only difference is the re-definition of $Q$.

\subsection{Application of the new methodology to the FX dataset}
\label{FX_reapplication}
%
%
%
%

\begin{figure}[H]
\centering
\includegraphics[width = 7cm, height = 6cm]{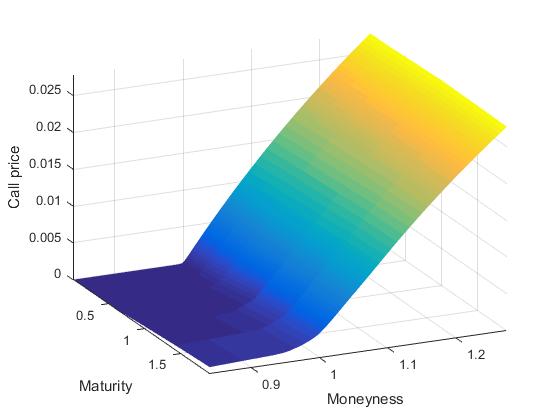}
\caption[Arbitrage-free call HKD/USD call-surface from 19/01/2016 obtained by using the structure-preserving $l_1$-recovery procedure for maturities ranging from one week to two years.]{The obtained call-surface on the HKD/USD from 19/01/2016 with maturities between one week and two years, using the structure-preserving $l_1$-recovery procedure.}
\label{FX_recov}
\end{figure}

When taking all the market maturities, with mollified bid- and mid- prices in the matrix $Q$ for a range of mollification parameters, one recovers the following surfaces of call prices and implied volatility (Figure \ref{FX_recov}). Again, the code is written in MATLAB, and we have used the GUROBI \cite{gurobi} optimiser to solve the linear program. Every recovered slice consists of $406$ strikes.

\begin{figure}[H]
\centering
\includegraphics[width = 14cm, height = 14cm]{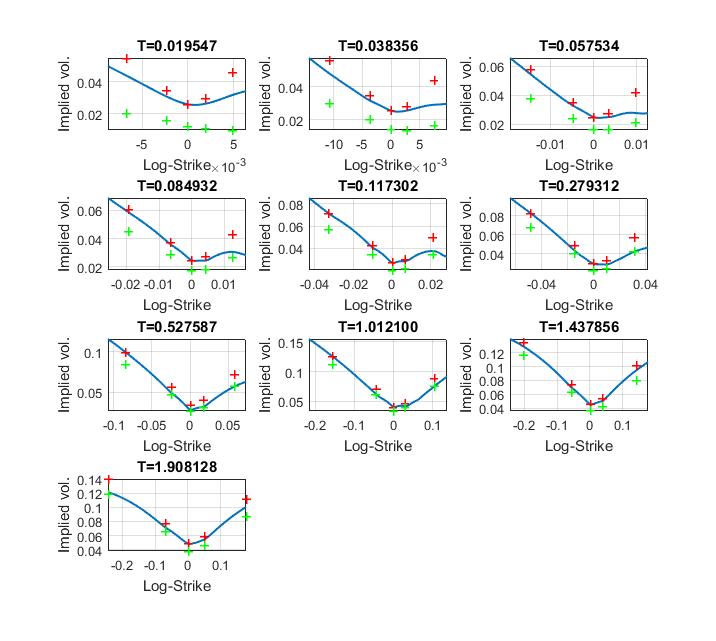}
\caption[Slice-per-slice plotting of the arbitrage-free implied volatility surface obtained from using the structure-preserving $l_1$-recovery procedure on HKD/USD bid-ask call prices (19. Jan 2016) for maturities ranging from one week to two years.]{The implied-volatility smiles as a function of the log-strike obtained by structure-preserving $l_1$-recovery.}
\label{slice_per_slice}
\end{figure}

We also provide the slice-per-slice plots of the above fit in Figure \ref{slice_per_slice}. Every subplot displays the implied-volatility smile with respect to the log-strike (the log of the strike/forward quotient), so that one can have a visual comparison with the Square-root SSVI fitting of Figure \ref{SVI_calibration}.

\begin{figure}[H]
\begin{center} 
\begin{align*}
\includegraphics[width = 7cm, height = 6cm]{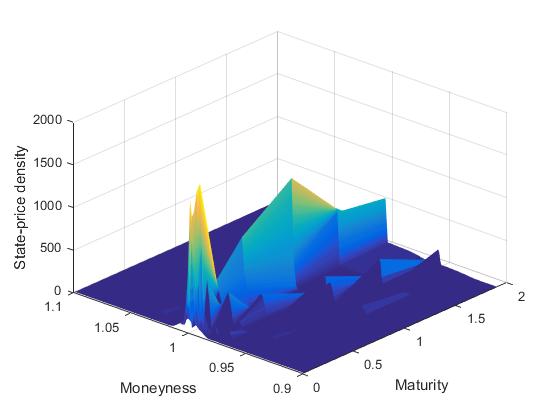}&\includegraphics[width = 7cm, height = 6cm]{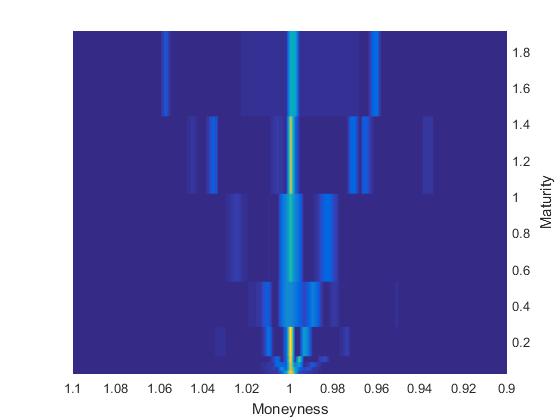}
\end{align*}
\caption[Density obtained from using the structure-preserving $l_1$-recovery procedure on HKD/USD bid-ask call prices (19. Jan 2016) for maturities ranging from one week to two years.]{The recovered state-price density surface of the HKD/USD. Right: the same surface, plotted from above. The presence of one main mode around the money becomes apparent. A second mode can be seen out of the money. A third, less marked mode appears in the money. Both secondary modes tend to drift away from the moneyness as time passes, due to the difference in interest rates in the US and Hong Kong.}
\label{FX_recov_density}
\end{center}
\end{figure}

We also inspect the state-price density; see Figure \ref{FX_recov_density}. Concerning the distribution, a few points can be noted. As usual, the highest mass is concentrated around the money. Moreover, the value of the density decays very rapidly around the money, which is consistent with the HKD currency being pegged.  Finally, one can notice that the distribution is not unimodal, but rather trimodal, with one secondary mode appearing out of the money, and a third, less pronounced, mode appears in the money. Note that the modes appear to ``drift away'' from the moneyness as the maturities expand. These observations conclude our study of this application.

\section{Conclusion}

In this article, we have developed a methodology to recover sparse call-option surfaces that are arbitrage free, match market prices up to desired precision, by showing that this problem could be set as a $l_1$-recovery problem, which itself could be recast as a linear program.

Applying this methodology to the S\&P 500 allowed us to recover the call surface, exhibiting a sparse decomposition and a smooth associated local-volatility surface.

When applying this methodology to the more challenging HKD/USD call-option dataset, the above described methodology failed and we had to take advantage of the problem structure so as to recover a finer option surface. The price to pay is a loss in the dynamics decomposition of the implied-volatility surface into a small number of explanatory functions, in favour of a recovery that relies more on consistent smoothing of the bid-ask spread.

\bibliographystyle{plain}
\bibliography{biblioCS}

\begin{thebibliography}{10}

\bibitem{C08}
E.~Cand{\`e}s.
\newblock The restricted isometry property and its implications for compressed
  sensing.
\newblock {\em Comptes Rendus Math{\'e}matiques}, 346(9):589--592, 2008.

\bibitem{CT05}
E.~Cand{\`e}s and T.~Tao.
\newblock Decoding by linear programming.
\newblock {\em Information Theory, IEEE Transactions on}, 51(12):4203--4215,
  2005.

\bibitem{CM05}
P.~Carr and D.~Madan.
\newblock A note on sufficient conditions for no arbitrage.
\newblock {\em Finance Research Letters}, 2(3):125--130, 2005.

\bibitem{CDS98}
S.~Chen, D.~Donoho, and M.~Saunders.
\newblock {Atomic Decomposition by Basis Pursuit}.
\newblock {\em SIAM journal on scientific computing}, 20(1):33--61, 1998.

\bibitem{C11}
I.J. Clark.
\newblock {\em Foreign Exchange Option Pricing: A Practitioner's Guide}.
\newblock The Wiley Finance Series. Wiley, 2011.

\bibitem{DDEK12}
M.~Davenport, M.~Duarte, Y.~Eldar, and G.~Kutyniok.
\newblock {Introduction to Compressed Sensing}.
\newblock In {\em {Compressed Sensing, Theory and Applications. Y. Eldar and G.
  Kutyniok, editors.}} Cambridge University Press, 2012.

\bibitem{DH07}
M.~Davis and D.~Hobson.
\newblock The range of traded option prices.
\newblock {\em Mathematical Finance}, 17(1):1--14, 2007.

\bibitem{DE03}
D.~Donoho and M.~Elad.
\newblock Optimally sparse representation in general (nonorthogonal)
  dictionaries via l1 minimization.
\newblock {\em Proceedings of the National Academy of Sciences},
  100(5):2197--2202, 2003.

\bibitem{E10}
M.~Elad.
\newblock {\em {Sparse and Redundant Representations:From Theory to
  Applications in Signal and Image Processing }}.
\newblock Springer-Verlag New-York, 2010.

\bibitem{F09}
M.~Fengler.
\newblock {Arbitrage-free smoothing of the implied volatility surface}.
\newblock {\em Quantitative Finance}, 9(4):417--428, 2009.

\bibitem{FH15}
M.~Fengler and L-Y Hin.
\newblock Semi-nonparametric estimation of the call-option price surface under
  strike and time-to-expiry no-arbitrage constraints.
\newblock {\em Journal of Econometrics}, 184(2):242--261, 2015.

\bibitem{GJ14}
J.~Gatheral and A.~Jacquier.
\newblock {Arbitrage-free SVI volatility surfaces}.
\newblock {\em Quantitative Finance}, 14(1):59--71, 2014.

\bibitem{G02}
M.~Ghomi.
\newblock The problem of optimal smoothing for convex functions.
\newblock {\em {Proceedings of the American Mathematical Society}},
  130(8):2255--2259, 2002.

\bibitem{gurobi}
Inc. Gurobi~Optimization.
\newblock Gurobi optimizer reference manual, 2015.

\bibitem{H97}
M.W. Hirsch.
\newblock {\em Differential Topology}.
\newblock Graduate Texts in Mathematics. Springer New York, 1997.

\bibitem{LY08}
D.~G. Luenberger and Y.~Ye.
\newblock {\em {Linear and Nonlinear Programming}}.
\newblock Springer US, 3rd edition, 2008.

\bibitem{N95}
B.~K. Natarajan.
\newblock Sparse approximate solutions to linear systems.
\newblock {\em SIAM journal on computing}, 24(2):227--234, 1995.

\bibitem{O12}
G.~Orosi.
\newblock {Empirical performance of a spline-based implied volatility surface}.
\newblock {\em Journal of Derivatives \& Hedge Funds}, 18(4):361--376, 2012.

\bibitem{RA15}
Holger Rauhut and Rachel Ward.
\newblock Interpolation via weighted $\ell_1$ minimization.
\newblock {\em Applied and Computational Harmonic Analysis}, 2015.

\bibitem{R10}
M.~Roper.
\newblock {Arbitrage Free Implied Volatility Surfaces}.
\newblock preprint, 2010.

\bibitem{R76}
W.~Rudin.
\newblock {\em Principles of Mathematical Analysis}.
\newblock International series in pure and applied mathematics. McGraw-Hill,
  3rd edition, 1976.

\end{thebibliography}
\end{document}